\documentclass{elsart}

\usepackage{graphicx}
\usepackage{amssymb}

\begin{document}

\begin{frontmatter}



\title{Heavy-heavy-light quark potential\\ in SU(3) lattice QCD}

\author[Kyoto]{Arata~Yamamoto \corauthref{cor}},
\corauth[cor]{Corresponding author.}
\ead{a-yamamoto@ruby.scphys.kyoto-u.ac.jp}
\author[Kyoto]{Hideo~Suganuma},
\author[YITP]{Hideaki~Iida}

\address[Kyoto]{Department of Physics, Faculty of Science, Kyoto University,\\Kitashirakawa, Sakyo, Kyoto 606-8502, Japan}
\address[YITP]{Yukawa Institute for Theoretical Physics, Kyoto University,\\Kitashirakawa, Sakyo, Kyoto 606-8502, Japan}
%\address[RCNP]{Research Center for Nuclear Physics, Osaka University, Mihogaoka 10-1, Osaka 567-0047, Japan}

\begin{abstract}
We perform the first study for the heavy-heavy-light quark ($QQq$) potential in SU(3) quenched lattice QCD with the Coulomb gauge.
The calculations are done with the standard gauge
and $O(a)$-improved Wilson fermion action on the $16^4$ lattice at $\beta=6.0$.
We calculate the energy of $QQq$ systems as the function of the distance $R$ between the two heavy quarks, and find that the $QQq$ potential is well described with a Coulomb plus linear potential form up to the intermediate distance $R \le$ 0.8 fm.
Compared to the static three-quark case, the effective string tension between the heavy quarks is significantly reduced by the finite-mass valence quark effect. 
This reduction is considered to be a general property for baryons.
\end{abstract}

\end{frontmatter}

\section{Introduction}
The inter-quark interaction is one of the fundamental and essential properties linking elementary physics and hadron physics.
In particular, the three-quark interaction in baryons is characteristic and complicated, reflecting the nontrivial gluonic dynamics based on the SU(3) gauge symmetry. 
In addition, the three-quark system has also large varieties of the quark motion, configuration, and so on.
The heavy-heavy-light quark ($QQq$) system is a suitable material to investigate such a light-quark effect on the three-quark interaction.

In 2002, the first doubly charmed baryon, $\Xi _{cc}^+(dcc)$, was experimentally observed at SELEX, Fermilab \cite{Ma02}.
In this experiment, a decay process $\Xi _{cc}^+ \to \Lambda _c^+ K^- \pi ^+$ was observed, and its mass was measured about 3519 MeV.
In another experiment, a decay process $\Xi _{cc}^+ \to p D^+ K^- $ was also confirmed \cite{Oc05}.
Doubly charmed baryons are also theoretically investigated in lattice QCD \cite{Le0102} and other approaches \cite{Ru75,Vi04}.
Since $c$ quark is much heavier than $d$ quark, this system can be idealized as the three-body system constructed with two static quarks and one finite-mass quark moving around.

Motivated by these considerations, we investigate the $QQq$ system in quenched lattice QCD.
We extract the $QQq$ potential $V_{QQq}(R)$, which is defined as the energy of the $QQq$ system in terms of the inter-heavy-quark distance $R$.
In lattice QCD, the quark-antiquark ($Q\bar Q$) potential \cite{Ba92} and the static three-quark ($3Q$) potential \cite{Ta0102} are already found to be described as the linear confinement potential plus the one-gluon-exchange Coulomb potential.
The confinement potential is proportional to the length of the gluonic flux connecting the quarks, and the string tension is about 0.89 GeV/fm.
In contrast to these static quark potentials, our $QQq$ potential includes not only the gluonic effect but also the nontrivial finite-mass valence quark effect.
Thus the $QQq$ potential behavior would have some difference from the static cases.

\section{Formalism}

\begin{figure}[t]
\begin{center}
\includegraphics[scale=0.4]{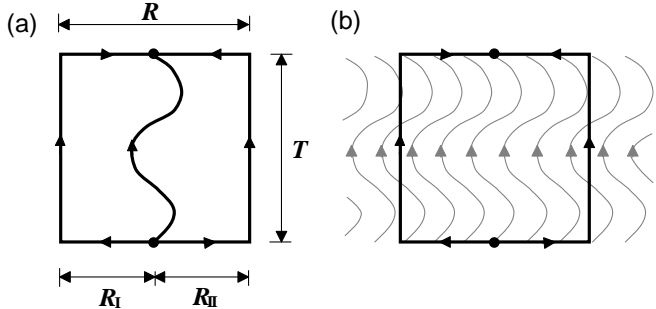}
\caption{\label{fig1}
(a)The gauge-invariant $QQq$ Wilson loop.
The wavy line represents the light-quark propagator and the straight line the heavy-quark trajectory.
(b)The ``wall-to-wall $QQq$ Wilson loop."
The gray wavy lines represent the wall-to-wall quark propagator, which propagates from the whole space at one time to that at another time.
}
\end{center}
\end{figure}

The $QQq$ Wilson loop is defined in almost the same manner as the static $3Q$ Wilson loop \cite{Ta0102}.
The difference is that the light quark is treated as the quark propagator $K^{-1}$.
The gauge-invariant $QQq$ Wilson loop with the spatial size $R$ and the temporal size $T$ is defined as
\begin{eqnarray}
W_{QQq} (R,T)\equiv \frac{1}{3!}\epsilon_{abc}\epsilon_{def} U^{\rm I}_{ad} U^{\rm II}_{be} K_{cf}^{-1}, 
\label{WQQq}
\end{eqnarray}
with the path-ordered product of the link variables $U^{k} = P\exp (ig\int_{\Gamma _k} dx^\mu A_\mu)$ along the heavy-quark trajectory $\Gamma _k \ (k={\rm I,II})$.
The subscripts $a,b,...,f$ are color indices.
The schematic figure is shown in Fig.~\ref{fig1}(a).
The $QQq$ potential is obtained as
\begin{eqnarray}
V_{QQq} (R)= - \lim _{T\rightarrow \infty} \frac{1}{T}\ln \langle W_{QQq}(R,T) \rangle.
\end{eqnarray}
The symbol $\langle \quad \rangle$ means the expectation value integrated over the gauge field.
All we have to do is to get the expectation value from lattice QCD for several values of $R$ and to give a suitable function form of $V_{QQq}(R)$.

\section{Simulation details}

\begin{table}[t]
\newcommand{\m}{\hphantom{$-$}}
\newcommand{\cc}[1]{\multicolumn{1}{c}{#1}}
\renewcommand{\tabcolsep}{0.5pc} 
\renewcommand{\arraystretch}{1} 
\caption{\label{tab1-1}
The lattice parameter $\beta =2N_c/g^2$, the corresponding lattice spacing $a$, the sweep numbers ($N_{\rm therm},N_{\rm sep}$) of the thermalization and separation for updating the gauge fields, the smearing parameters ($\alpha,N_{\rm smr}$), and the clover coefficient $c$.
}
\begin{center}
\begin{tabular}{cccccccc}
\hline\hline
$\beta$ & $a$ [fm] & lattice size & $N_{\rm therm}$ &
$N_{\rm sep}$ & $\alpha$ & $N_{\rm smr}$ & $c$ \\
\hline
6.0 & 0.10 & $16^4$ & 10000 & 500 & 2.3 & 40 & 1.479\\
\hline\hline
\end{tabular}
\end{center}
%\end{table}

%\begin{table}[!h]
\caption{\label{tab1-2}
The correspondence between $\kappa$ and the used gauge configuration number $N_{\rm conf}$.
The list shows the pion mass $m_{\pi}$, the $\rho$ meson mass $m_{\rho}$, and the approximate constituent quark mass $M_q\simeq m_{\rho}/2$.
The meson masses are obtained from the meson correlator with the wall source and the point sink.
The statistical error is estimated with the jackknife method.
}
\begin{center}
\begin{tabular}{ccccc}
\hline\hline
$\kappa$ & $N_{\rm conf}$ & $m_{\pi}a$ & $m_{\rho}a$ & $M_q$ \\
\hline
0.1200 & 1000 & 1.446(1) & 1.472(2) &  1.5 GeV\\
0.1300 &  300 & 0.900(2) & 0.949(1) &  1 GeV\\
0.1340 &  300 & 0.643(1)  & 0.716(1) &  700 MeV\\
0.1380 & 1000 & 0.304(1) & 0.467(2)&  500 MeV\\
\hline\hline
\end{tabular}
\end{center}
\end{table}

We generate the SU(3) gauge configurations with $\beta=6.0$ and $16^4$ isotropic lattice at the quenched level.
We adopt the standard plaquette gauge action, and the pseudo-heat-bath algorithm to update the gauge field.
Periodic boundary conditions are imposed on the space-time boundaries. 
We apply the smearing method in Refs.~\cite{Ta0102,Bo92} to the spatial link variables of the $QQq$ Wilson loop. 
The smearing method changes a stringy link to a spatially-extended flux tube, and enhances the ground-state component without changing the physical content.  
The method has two parameters, a real parameter $\alpha$ and the iteration number $N_{\rm smr}$, and our choice of $\alpha$ and $N_{\rm smr}$ is based on the static $3Q$ case \cite{Ta0102}.
These simulation parameters are summarized in Table \ref{tab1-1}.
The lattice spacing $a\simeq 0.10$ fm is determined so as to reproduce the string tension of the $Q\bar Q$ potential to be 0.89 GeV/fm .
We use the lattice unit for most part of the paper.

For the light-quark propagator, we adopt the clover fermion action, which is the $O(a)$-improved Wilson fermion action \cite{El97}.
The clover coefficient $c$ in this action is given from the mean field value $u_0$ of the link variable for the tadpole improvement.
We determine $c$ and $u_0$ from the ensemble average of the all plaquette values $P_{\mu\nu}(n)$ as
$c=1/u_0^3$ and $u_0=\langle \sum_n \sum_{\mu>\nu} \frac{1}{3}{\rm ReTr} P_{\mu\nu}(n) \rangle ^{1/4}$.
The measured value of $u_0$ is 0.87779(2) in our case.
To investigate the light-quark-mass dependence, we take different four light-quark hopping parameters, $\kappa =0.1200$, 0.1300, 0.1340, and 0.1380.
Their correspondences to the light-quark masses are shown in Table \ref{tab1-2}.
The constituent quark mass $M_q$ is roughly estimated with the half of the $\rho$ meson mass.

\begin{figure}[t]
\begin{center}
\includegraphics[scale=1]{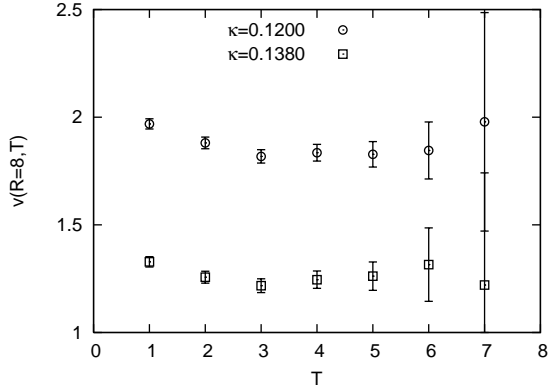}
\caption{\label{fig2}
The typical example of the effective mass plot: $v(R=8,T)$ with $(R_{\rm I},R_{\rm II})=(4,4)$.
The upper data correspond to the heaviest case, $\kappa =0.1200$, and the lower data the lightest case, $\kappa=0.1380$.
All the scales are measured in lattice unit.
}
\end{center}
\end{figure}

In Eq.~(\ref{WQQq}) and Fig.~\ref{fig1}(a), the $QQq$ Wilson loop is defined as a single gauge invariant loop.
To reduce the statistical error, we adopt the following prescription.
The light-quark propagator is spatially extended as the wall source and the wall sink.
(The ``wall" means the average over all spatial sites.)
This propagator is averaged over all the spatial sites $n_{\rm src}$ of the source and all the spatial sites $n_{\rm sink}$ of the sink at the fixed time separation $T$, as $K^{-1}_{\rm wall}(T)\propto \sum_{n_{\rm src}}\sum_{n_{\rm sink}} K^{-1}(n_{\rm src},n_{\rm sink},T)$.
The ``wall-to-wall $QQq$ Wilson loop" is constructed from this wall-to-wall quark propagator and the heavy-quark trajectories, and its schematic figure is depicted in Fig.~\ref{fig1}(b).
Because such a propagator is independent of the spatial position, we can easily sum up the parallel translated wall-to-wall $QQq$ Wilson loops in the whole space.
This summing up drastically suppresses the statistical error, owing to the large statistics, e.g., $16^3$ times larger in our case.
In the gauge invariant formalism, only gauge invariant components in wall-to-wall $QQq$ Wilson loops remain.

When we calculate the $QQq$ potential in the gauge invariant way, we find that the statistical and systematic errors are severely large, especially for lighter quark mass case.
For the error reduction, we fix the gauge configurations with the Coulomb gauge.
We should note that, in the wall-to-wall $QQq$ Wilson loops with the Coulomb gauge, gauge variant components also remain due to the nonlocal nature of the gauge.
However, the Coulomb gauge fixing empirically does not affect the long-range physics, such as the string tension \cite{Gr03}.
In this paper, we are mainly interested in the long-range behavior of the $QQq$ potential, and investigate it with the Coulomb gauge.
We have also calculated in the Landau gauge, and obtained the same result as the Coulomb gauge.
 
\section{Results}

\begin{table}[t]
\begin{center}
\newcommand{\m}{\hphantom{$-$}}
\newcommand{\cc}[1]{\multicolumn{1}{c}{#1}}
\renewcommand{\tabcolsep}{0.5pc} 
\renewcommand{\arraystretch}{1} 
\caption{\label{tab2}
The lattice QCD results for the Coulomb-gauge-fixed $QQq$ potential $V_{QQq}$ at $\kappa =0.1380$. 
$R$ and $(R_{\rm I},R_{\rm II})$ denote the loop size defined in Fig.~\ref{fig1}.
The results with different fit ranges of $T$ are also shown.
All the values are in lattice unit, and the statistical error is estimated with the jackknife method.
}
\begin{tabular}{cccc}
\hline\hline
$R$ & $(R_{\rm I},R_{\rm II})$ & $V_{QQq} \ {\rm in}\ T=[4,8]$& $V_{QQq} \ {\rm in} \ T=[5,8]$\\
\hline
1 & (0,1) & 0.877(2) & 0.873(2)\\
2 & (0,2) & 0.971(7) & 0.959(9)\\
  & (1,1) & 0.969(8) & 0.958(10)\\
3 & (0,3) & 1.047(4) & 1.045(7)\\
  & (1,2) & 1.045(4) & 1.043(8)\\
4 & (0,4) & 1.083(11) & 1.067(17)\\
  & (1,3) & 1.079(10) & 1.063(16)\\
  & (2,2) & 1.078(10) & 1.063(15)\\
5 & (0,5) & 1.136(6) & 1.122(3)\\
  & (1,4) & 1.131(6) & 1.117(4)\\
  & (2,3) & 1.130(6) & 1.116(5)\\
6 & (0,6) & 1.170(13) & 1.151(24)\\
  & (2,4) & 1.157(16) & 1.136(30)\\
  & (3,3) & 1.157(16) & 1.136(31)\\
7 & (0,7) & 1.219(21) & 1.220(50)\\
  & (3,4) & 1.207(24) & 1.209(60)\\
8 & (0,8) & 1.262(11) & 1.283(21)\\
  & (4,4) & 1.255(6) & 1.271(10)\\
\hline\hline
\end{tabular}
\end{center}
\end{table}

To decide the fit range of $T$, we define the effective mass
\begin{eqnarray}
v(R,T)\equiv \ln \frac{\langle W_{QQq}(R,T) \rangle}{\langle W_{QQq}(R,T+1)\rangle},
\end{eqnarray}
and seek its plateau region against $T$.
If the state is dominated by a single component, $v(R,T)$ is independent of $T$.
Typical cases are plotted in Fig.~\ref{fig2}.
All the statistical errors are estimated with the jackknife method here and below.
We see that, in $T \ge 3$, the effective mass is approximately flat and thus the ground state component dominates.
By fitting $\langle W_{QQq} \rangle$ with a single exponential form $Ce^{-V_{QQq}T}$, we obtain the $QQq$ potential values $V_{QQq}$ of different loop sizes, as partially listed in Table \ref{tab2}.
As mentioned above, $V_{QQq}$ is almost unchanged for the different fit range of $T$.

Remarkably, $V_{QQq}$ depends only on $R(=R_{\rm I}+R_{\rm II})$ and not on the combination of ($R_{\rm I},R_{\rm II}$). 
This property is suitable for the potential calculation for the following reason.
If the fit range of $T$ is large enough, the expectation value of the Wilson loop is dominated by the ground-state component, and does not depend on its condition of the source and sink.
Then the $QQq$ potential does not depend on ($R_{\rm I},R_{\rm II}$) or more arbitrary choice of the junction points.

\begin{figure}[t]
\begin{center}
\includegraphics[scale=1.1]{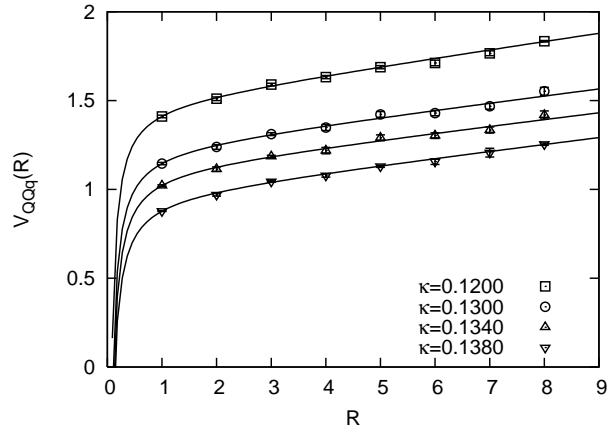}
\caption{\label{fig3}
The lattice QCD data of the Coulomb-gauge-fixed $QQq$ potential $V_{QQq}$ plotted against the inter-heavy-quark distance $R$ for the different four values of $\kappa$.
The solid curves are the best-fit functions of Eq.~(\ref{VQQq}).
All the scales are measured in lattice unit.
}
\end{center}
\end{figure}

To obtain the functional form of $V_{QQq}(R)$, we consider the fit function,
\begin{eqnarray}
V_{QQq}(R)=\sigma _{\rm eff}R-\frac{A_{\rm eff}}{R}+C_{\rm eff},
\label{VQQq}
\end{eqnarray}
as the analogy of the $Q\bar Q$ potential.
The subscript ``eff" means the effective values including the light-quark effect.
We find that this function is fairly suitable for $V_{QQq}$, at least in this calculated region of $R$, i.e., $R\le0.8$ fm.
The best-fit parameters are shown in Table \ref{tab3}, and the corresponding potential form is depicted in Fig.~\ref{fig3}.
In the static $Q\bar Q$ and $3Q$ potentials, the string tension and the Coulomb coefficient are obtained as
\begin{eqnarray}
\sigma _{Q\bar Q} \simeq  \sigma _{3Q} \simeq  0.045, \quad
\frac{1}{2}A_{Q\bar Q} \simeq  A_{3Q} \simeq  0.13
\end{eqnarray}
in the lattice unit at $\beta =6.0$ \cite{Ta0102}.
The effective Coulomb coefficient $A_{\rm eff}$ is almost the same value as $A_{3Q}$.
As for the effective string tension $\sigma _{\rm eff}$, in the heaviest valence quark case, $\sigma _{\rm eff}$ does not deviate from $\sigma_{3Q}$ within the statistical error.
However, $\sigma _{\rm eff}$ obtained in the lightest quark case is significantly reduced from $\sigma_{3Q}$,
\begin{eqnarray}
\sigma_{\rm eff} < \sigma_{3Q},
\end{eqnarray}
provided that the finite volume effect is negligible.
Therefore, when the valence quark is light, its effect reduces the effective string tension $\sigma _{\rm eff}$ from the string tension $\sigma_{3Q}$ in the static $3Q$ case. 

\begin{table}[b]
\newcommand{\m}{\hphantom{$-$}}
\newcommand{\cc}[1]{\multicolumn{1}{c}{#1}}
\renewcommand{\tabcolsep}{0.5pc} 
\renewcommand{\arraystretch}{1} 
\caption{\label{tab3}
The best-fit values of $\sigma_{\rm eff}$, $A_{\rm eff}$, and $C_{\rm eff}$ in Eq.~(\ref{VQQq}), and their $\chi ^2$ over the degree of freedom $N_{\rm dof}$.
All the values are in unit of the lattice spacing $a\simeq 0.10$ fm.
}
\begin{center}
\begin{tabular}{ccccc}
\hline\hline
$\kappa$ & $\sigma_{\rm eff}$ & $A_{\rm eff}$ & $C_{\rm eff}$ & $\chi ^2/N_{\rm dof}$\\
\hline
0.1200 & 0.045(2) & 0.12(2) & 1.49(2) & 1.31\\
0.1300 & 0.038(4) & 0.13(2) & 1.23(3) & 1.18\\
0.1340 & 0.037(4) & 0.13(2) & 1.12(2) & 1.11\\
0.1380 & 0.037(2) & 0.13(1) & 0.97(1) & 1.16\\
\hline\hline
\end{tabular}
\end{center}
\end{table}

\section{Discussion and summary}

In our $QQq$ potential, the significant result is the reduction of the effective string tension.
To understand this result, we compare the string tension with the effective string tension.
In the $Q\bar Q$ case, the color flux-tube length corresponds to the distance between the quark and the antiquark, and thus $\sigma _{Q\bar Q}$ is the proportionality coefficient of the inter-quark distance in the confinement potential.
Similarly, in the $3Q$ case, $\sigma _{3Q}$ is the proportionality coefficient of the flux-tube length that minimally connects the three quarks \cite{Ta0102,Ic03,Co04}.
However, in the $QQq$ system, the inter-quark distance $R$ is {\it not} the flux-tube length.
The $QQq$ flux-tube length is determined by complicated light-quark dynamics, and therefore $V_{QQq}(R)$ does not have to be proportional to $R$.
The actual lattice QCD results suggest that the long-range $QQq$ potential is proportional to $R$ at least for $R\le 0.8$ fm, but the effective string tension $\sigma_{\rm eff}$ as its proportionality coefficient is reduced by the existence of the light quark, which is conjectured to originate from the difference between $R$ and the flux-tube length.

Now we consider the asymptotic behavior of the $QQq$ potential.
In the large $R$ limit, the confinement potential between the two heavy quarks will dominate, and the color flux-tube length approximately equals to $R$.
Therefore $\sigma_{\rm eff}$ would approach to $\sigma_{3Q}$, which will be confirmed with a larger-volume calculation.
There is another asymptotic behavior we can check, that is, the limit in which the light quark is ``heavy".
If the light-quark mass goes to infinite, the $QQq$ system should correspond to the static $3Q$ system.
It means that $\sigma _{\rm eff}$ becomes larger in the larger light-quark mass and $\sigma _{\rm eff} = \sigma _{3Q}$ in the infinite mass limit.
We can confirm this behavior from the lattice result in the heaviest case, $\kappa =0.1200$, in Table \ref{tab3}.
 
In our another work \cite{YaUP}, we investigate the same $QQq$ potential in a non-relativistic quark model.
It reproduces the lattice QCD result, and enables us to understand the reduction mechanism of $\sigma_{\rm eff}$.
We confirm that a geometrical difference between the flux-tube length and $R$ is essential for the reduction of $\sigma_{\rm eff}$, as conjectured above. 

We mention another possible light-quark effect on the interquark potential, i.e., the sea quark effect.
The sea quark effect causes the string breaking, which is the disappearance of the string tension at a long distance.
The sea quark effect is obtained with the unquenched lattice QCD calculation, and is found to be important at the long range where the flux-tube length is larger than about 1 fm \cite{Ba05}.
The effective string tension in the realistic $QQq$ system would be also affected by such a sea quark effect in further large-$R$ region.

Although our lattice calculation is restricted to the specific gauge, we come to the following general conclusion about the effective string tension, which would not be affected by the nonlocality of the Coulomb gauge.
Owing to the existence of a mobile quark, the effective string tension $\sigma_{\rm eff}$ between the other two quarks can be reduced from the string tension in mesons  $\sigma_{Q\bar Q}$ or baryons $\sigma_{3Q}$. 
Since the reason for the reduction is fairly simple and general, this argument holds for not only $QQq$ systems but also three finite-mass quark systems, i.e., ordinary baryons, such as a nucleon.
In addition, this can be also applied to the multi-quark system including light quarks \cite{Gr96}.
For more quantitative calculation for realistic hadrons, we need careful consideration about realistic quark masses, finite-volume effects, sea quark effects, and more complicated valence quark effects. 

In summary, we have studied the $QQq$ potential in SU(3) quenched lattice QCD.
For the error reduction, we have fixed the gauge field with the Coulomb gauge, and investigated the long-distance behavior.
The effective string tension $\sigma _{\rm eff}$ is  10-20\% reduced, compared to the string tension in the static case, in $R \le 0.8$ fm and 0.5 GeV $\le M_q \le 1$ GeV.
The light-quark mass dependence of the reduction is also investigated.
The effective string tension means the inter-two-quark confining force in baryons, and its reduction by the finite-mass valence quark effect is conjectured to be a general property for baryons and multi-quark hadrons.
The significant change of the fundamental inter-quark force is important not only for QCD but also for the quark-hadron and nuclear physics.

\section*{Acknowledgements}
We thank Dr.~T.~T.~Takahashi, Dr.~T.~Umeda and Prof.~S.~J.~Brodsky for useful comments and discussions.
H.~S.~was supported by a Grant for Scientific Research [(C) No.19540287] in Japan.
The lattice QCD calculations are done on NEC SX-8R at Osaka University.


\begin{thebibliography}{99}
\bibitem{Ma02} M. Mattson {\it et al.} (SELEX Collaboration), Phys. Rev. Lett. {\bf 89} (2002) 112001.
\bibitem{Oc05} A. Ocherashvili {\it et al.} (SELEX Collaboration), Phys. Lett. {\bf B628} (2005) 18.
\bibitem{Le0102} R. Lewis, N. Mathur, and R. M. Woloshyn, Phys. Rev. D {\bf 64} (2001) 094509;  N. Mathur, R. Lewis, and R. M. Woloshyn, Phys. Rev. D {\bf 66} (2002) 014502.
\bibitem{Ru75} A. De Rujula, H. Georgi, and S. Glashow, Phys. Rev. D {\bf 12} (1975) 147.
\bibitem{Vi04} J. Vijande, H. Garcilazo, A. Valcarce, and F. Fernandez, Phys. Rev. D {\bf 70} (2004) 054022.
\bibitem{Ba92} G. S. Bali and K. Schilling, Phys. Rev. D {\bf 46} (1992) 2636.
\bibitem{Ta0102} T. T. Takahashi, H. Matsufuru, Y. Nemoto, and H. Suganuma, 
Phys. Rev. Lett. {\bf 86} (2001) 18; Phys. Rev. D {\bf 65} (2002) 114509; F. Okiharu, H. Suganuma, and T. T. Takahashi, Phys. Rev. D {\bf 72} (2005) 014505; Phys. Rev. Lett. {\bf 94} (2005) 192001.
\bibitem{Bo92} S. P. Booth {\it et al.} (UKQCD Collaboration), Phys. Lett. {\bf B275} (1992) 424.
\bibitem{El97} A. X. El-Khadra, A. S. Kronfeld, and P. B. Mackenzie, Phys. Rev. D {\bf 55} (1997) 3933;
G. P. Lepage and P. B. Mackenzie, Phys. Rev. D {\bf 48} (1993) 2250.
\bibitem{Gr03} J. Greensite and S. Olejnik, Phys. Rev. D {\bf 67} (2003) 094503.
\bibitem{Ic03} H. Ichie, V. Bornyakov, T. Streuer, and G. Schierholz, Nucl. Phys. {\bf A721} (2003) C899; V. G. Bornyakov, H. Ichie, Y. Mori, D. Pleiter, M. I. Polikarpov, G. Schierholz, T. Streuer, H. Stuben, and T. Suzuki, Phys. Rev. D {\bf 70} (2004) 054506.
\bibitem{Co04} J. M. Cornwall, Phys. Rev. D {\bf 69} (2004) 065013.
\bibitem{YaUP} A. Yamamoto and H. Suganuma, Phys. Rev. D {\bf 77} (2008) 014036. 
\bibitem{Ba05} G. S. Bali, H. Neff, T. D\"{u}ssel, T. Lippert, and K. Schilling, Phys. Rev. D {\bf 71} (2005) 114513. 
\bibitem{Gr96} A. M. Green, J. Lukkarinen, P. Pennanen, and C. Michael, Phys. Rev. D {\bf 53} (1996) 261; P. Pennanen, A. M. Green, and C. Michael, Phys. Rev. D {\bf 59} (1999) 014504.
\end{thebibliography}
\end{document}